\documentstyle[twoside,12pt,draft]{article}
\begin{document}
\input epsf
%
\normalsize
\par
\begin{flushright}
ISN Grenoble report 01-13
\end{flushright}
\par
\vspace{1cm}
\begin{center}
{\Large \bf Secondary electrons and positrons in near earth orbit}
\vspace{0.3cm}
\par
   L. Derome, M. Bu\'enerd, and Yong Liu
\par
\vspace{0.2cm}
{\sl Institut des Sciences Nucl\'{e}aires, IN2P3/CNRS, 
53 av. des Martyrs, 38026 Grenoble cedex, France}
\par
\normalsize
\end{center}
\vspace{0.5cm}
\begin{center}
\parbox{16cm}{\underline{Abstract}: 
The secondary e$^+$ and e$^-$ populations in near earth's orbit have been calculated by 
simulation. The results are in very good agreement with the recent AMS measurements. 
The e$^+$ over e$^-$ flux ratio for particles below the geomagnetic cutoff appears to 
be due to the geomagnetic East-West effect.}
\end{center}
\setcounter{page}{1}
The study of particle populations in the earth environment has been a field of intense 
research activity for more than 50 years \cite{ALB}. After a period of latence,
the field is regaining interest with the new generation of ambitious  
experimental projects. They should bring major improvements to the instrumental accuracy of the measurements and orders of magnitudes to the statistics of counting, with respect to previous experiments (see the references quoted in \cite{PROT,LEPT} for the current and historical context). Some new measurements of particle flux close to earth have been performed recently by the AMS experiment, providing a large sample of new data. 
These measurements open new prospects for high precision studies of the cosmic ray 
(CR)-atmosphere interactions and the dynamics of particles in the magnetosphere. 
This is of particular importance for the atmospheric neutrino issue, in the current context of this research field, since the knowledge of the (electron and muon) neutrino flux produced in the same decay chain of pions in the atmosphere is necessary for interpreting the results of the underground neutrino experiments \cite{NEUTRINOS}. Consequently, a good understanding of the electron and positron flux is highly relevant. 
\par
The proton distributions measured below the earth geomagnetic cutoff (GC) by AMS \cite{PROT} 
have been successfully interpreted recently in terms of interactions of the (proton) cosmic 
ray (CR) flux with the atmosphere \cite{PAP1}. This work is referred to as I in the following. 
Extending the work reported in I on protons, the present paper addresses the interpretation 
of the flux of positrons and electrons measured by AMS below the cutoff (subGC) \cite{LEPT} 
in the same phenomenological framework. 
\par
\begin{figure}[htbp]
\begin{center}
\vspace{1cm}
\hspace{-1cm}
\epsfysize=10cm
\epsfbox{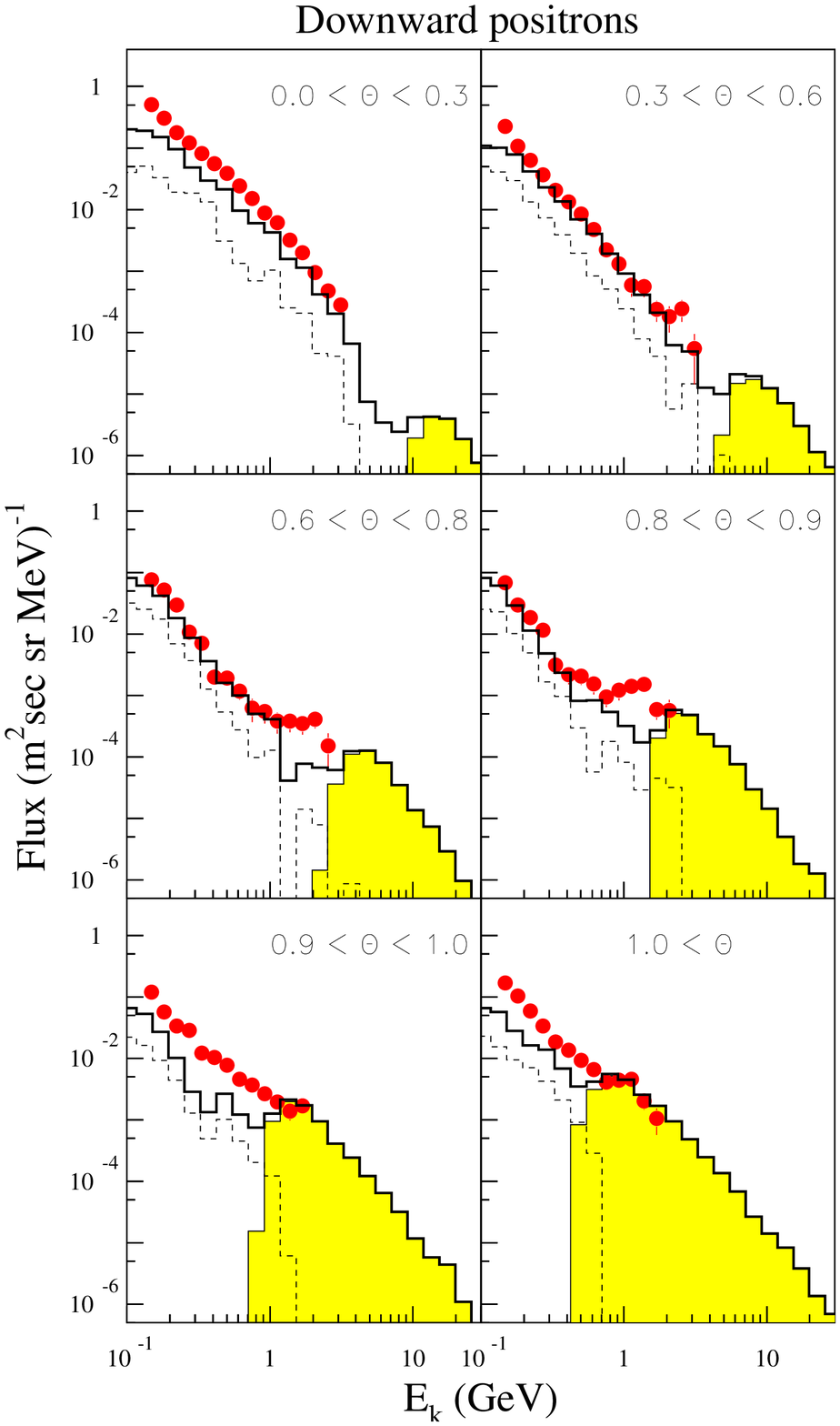} 
\epsfysize=10cm
\epsfbox{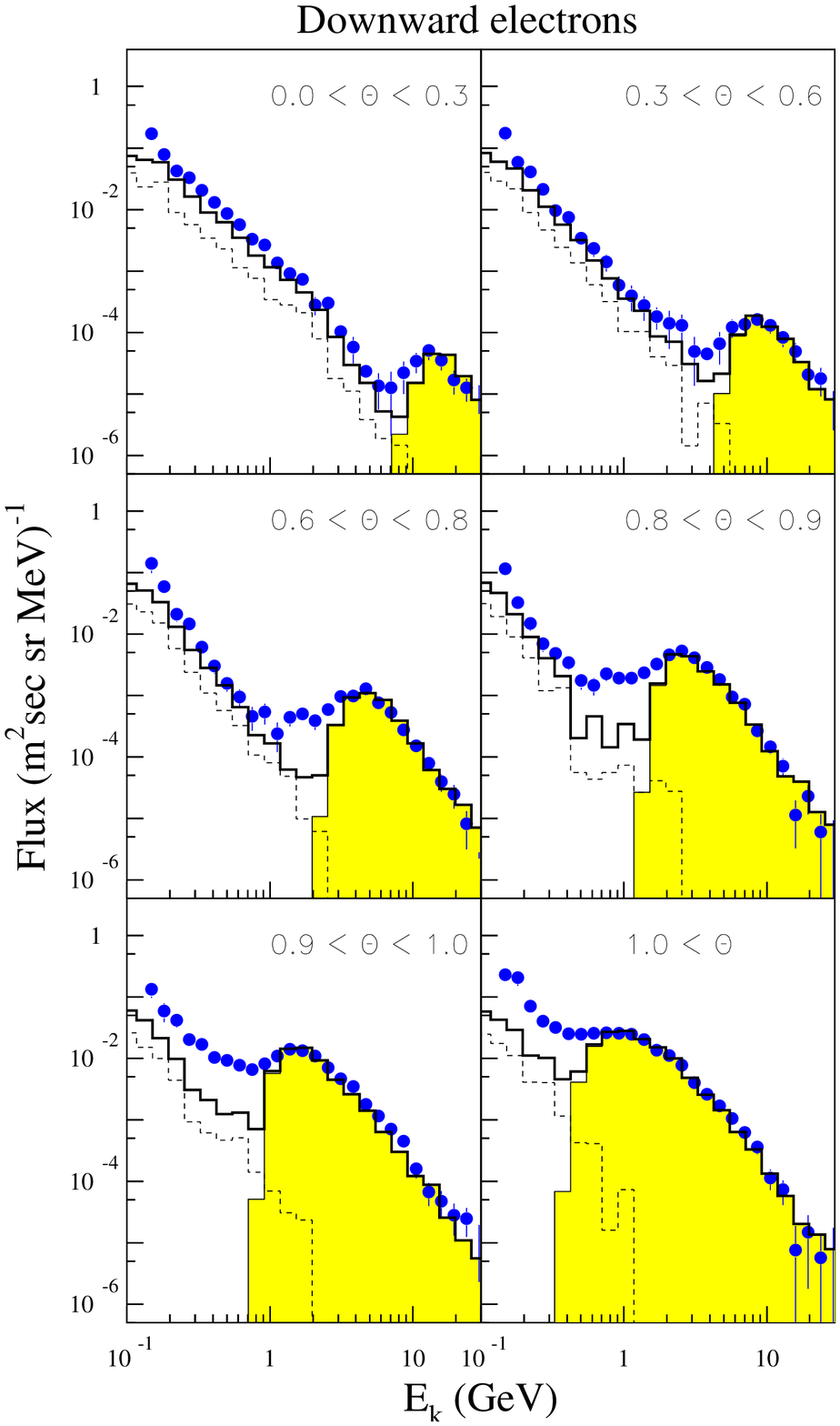}
\end{center}
\vspace{-0.5cm}
\caption{\small\it Simulated kinetic energy spectra for $e^+$ and $e^-$ (histograms), 
compared to the AMS data (full circles) in bins of geomagnetic latitude. The histograms 
correspond to the primary cosmic (shaded), and total (thick solid) flux respectively. 
The dashed histograms show the contribution of pair conversion from gammas originating 
from neutral-pion decay and electron/positron Bremstrahlung.}
\label{SPEC}
\end{figure}

The lepton distributions observed by AMS showed a few remarkable features, of which the main two follow. First, the kinetic energy spectra have a strong subGC (then 
secondary) component (figure~\ref{SPEC}), similar to those observed in the proton flux. 
Second, the ratio of the e$^{+}$ over e$^{-}$ flux is large in the equatorial region 
($\approx$~4) and decreasing towards higher latitudes ($\approx$~1). The measured 
spectra extended from the kinetic energy threshold of the spectrometer at $\approx$0.15~GeV, 
up to about 30~GeV for e$^{-}$ and 3~GeV for e$^{+}$. At low energy, the subGC secondary 
spectral yield decreases rapidly from a maximum around the low energy limit of the spectrometer 
up to the GC energy, around 15~GeV/c for equatorial latitudes (electrons) where the high 
energy tail of the subGC component merges with the galactic CR spectrum above GC (figure 
\ref{SPEC}). Interestingly, the subGC lepton and proton distributions at low 
latitudes have strikingly similar shapes \cite{PROT,LEPT}. This indicates that the two  
populations could also have similar dynamical origins, providing a further motivation to 
investigate the lepton populations along the same lines as in I.
\par
The observed subGC leptons are expected to originate mainly from pion production. The decay 
of charged pions produced in $p+A\rightarrow\pi + X$ ($A$ atmospheric nuclei) collisions is 
expected to be dominant. The pair conversion of gammas, either decaying from neutral pions 
produced in the same reactions or radiated by electron Bremstrahlung could also provide a 
significant contribution. The observed charge asymmetry of the electron-positron yield could 
be induced by the well known East-West (EW) geomagnetic effect \cite{EW} due to the earth 
magnetic (dipole) field. The latter consists of an EW angle dependence of the GC momentum,
which is maximum at one side and minimum at the opposite for a given charge of the particle, 
and conversely for the other charge (see below). The difference of production cross sections 
for $\pi^+$ and $\pi^-$, which is large at low incident proton energies \cite{COCH} 
(see \cite{PICS} for $pp\rightarrow\pi+X$ asymmetry), could also generate the observed effect.
It could also result from a combination of both effects.
\par
The study undertaken with the above ideas in mind was based on the same Monte-Carlo 
simulation program for the lepton production and propagation as used for protons in I 
(see this reference for details) \footnote{While this paper was being refereed, a previous 
analytic approach to this problem has been pointed out to the authors \cite{GR77}. See also 
\cite{VO94} for a Monte-Carlo approach of 20-200~MeV electron and positron flux.}. For 
the present study, the pion production cross sections in $p+A$ collisions, the subsequent 
decays $\pi\rightarrow\mu\rightarrow e$ and  $\pi_0\rightarrow\gamma\gamma$, lepton 
Bremstrahlung and pair conversion cross sections, have been included in the 
computation program together with some technical improvements in the event processing. 
Incident CRs ($p, e^+, e^-$) are generated according to their natural abundance using 
the recent AMS measurements \cite{LEPT,COSP}. They are propagated in the earth magnetic 
field and atmosphere and are allowed to interact with atmospheric nuclei. 
Each interaction can produce nucleons and pions according to their respective production 
cross sections and multiplicities. Each produced particle is then processed the same way, 
leading to the possible development of atmospheric cascades in which each particle history 
is traced and recorded. The e$^+$ and e$^-$ populations are generated by counting particles 
each time they cross upward or downward the mean altitude of AMS (370~km) within the 
detector acceptance (upwards and downwards particles are equivalent to splash-Albedo and 
secondary plus reentrant+Albedo particles in the Geophysics terminology, respectively).
\par
The pion production cross-sections on atmospheric nuclei are a critical input to the 
calculations since they are expected to govern the observed lepton populations. For this 
reason, the event generator has been built to provide as accurately as possible these cross 
sections over the range of sensitivity of the measurements. The latter extends from a few 
hundred MeV/c above threshold on the low energy side, up to a few hundred GeV/c at high 
energy, where the sensitivity vanishes with the spectral distribution of the cosmic flux 
which varies like $\approx E^{-2.7}$ \cite{SI83,COSP}. The maximum of sensitivity is 
expected around 10~GeV. A set of experimental angular distributions of inclusive cross 
sections for proton induced charged pion production, measured around this latter energy 
\cite{CHO,ABBO} could be well fitted by means of available parametrizations \cite{KALI,MOKH}. 
However, the very broad incident energy range to be covered here required the functional 
form used, to be modified. Good results could be obtained between 0.73~GeV and 200~GeV 
incident kinetic energies \cite{YONG} using the data from \cite{COCH,CHO,ABBO,PIDAT}. A set
of lower energy data were used \cite{LPIDAT} to constrain the energy dependence of the 
integrated cross section close to threshold. The $\pi^0$ production cross section used was 
the mean value of $\pi^+$ and $\pi^-$ cross sections, in agreement with \cite{PI0} since
no detailed data could be found.
\par
The simulation program was run for 2$\cdot$10$^7$ events reaching the earth. Note that statistics 
is limited by the very computer-time consuming processing of trapped low energy particles. 
The proton distributions obtained here (not shown) are in similar, even better at low 
energy, agreement with the data as reported in I. Figure~\ref{SPEC} shows the $e^+$ and 
$e^-$ downward spectra measured by AMS \cite{LEPT} compared to the results of the 
simulation, below and above GC, for the same geomagnetic latitudes. The two subGC lepton 
populations are seen to be very well reproduced in the equatorial region as well as at 
intermediate latitudes. However, in the polar region the simulated flux somewhat underestimates 
the experimental flux of secondaries. Note that at this latitude where the GC is very low, the 
latter is a most difficult spectrum to account for, since the cosmic proton energy range 
covers the whole domain from the threshold region up to the upper energy limit mentionned above. 
The neutral pion contributions are also shown on the figure. They have the same shape as for 
charged pions, and their overall contribution amounts to 17\%(22\%) of the total 
positron(electron) yield, whereas the Bremstrahlung contribution is 9\%(15\%). Note that 
the calculated distributions are entirely determined by the physics input of the simulation 
without any adjustable parameter. A similar agreement is obtained for upward $e^+$ and $e^-$ 
spectra (not shown).  As it was found for protons in I, the low energy flux obtained for 
$e^+$ and $e^-$ in the equatorial region is mainly due to trapped particles, which are in 
fact produced with a rather low probability. Their large contribution is due essentially to 
their high crossing multiplicity of the detection altitude. Long simulation runs are required 
for significant statistics of such events to be produced.
%
\par
\begin{figure}[htbp]
\begin{center}
\vspace{-1cm}
\hspace{-2cm}
\epsfysize=15cm
\epsfbox{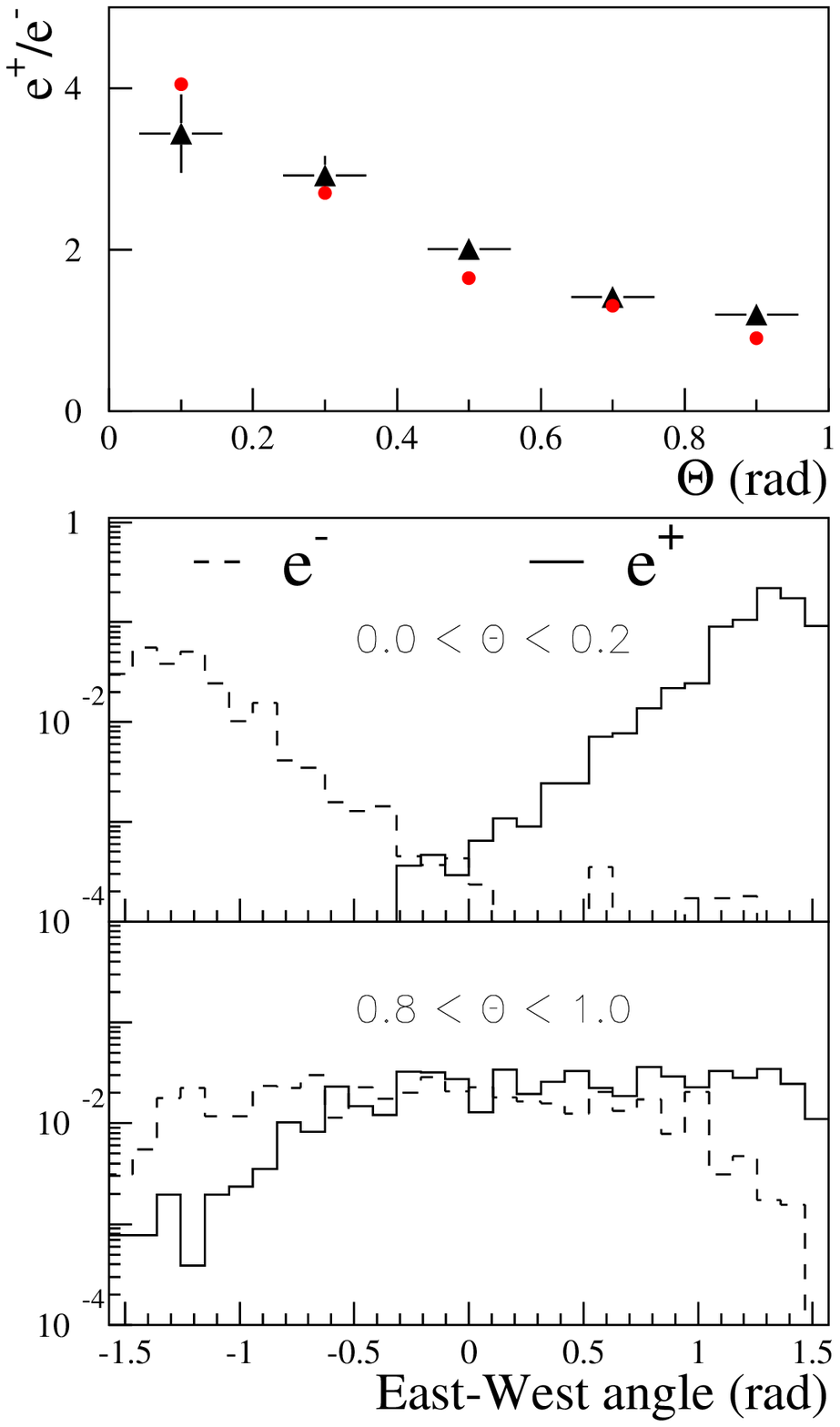} \\
\end{center}
\vspace{0.5cm}
\caption{\small\it Top:  e$^+$/e$^-$ flux ratio obtained from this work (triangles), 
compared to the AMS data (full circle) as a function of the geomagnetic latitude. Middle: 
EW angle distributions of the incident protons having generated a secondary lepton at the 
interaction point, for the equatorial latitude bin. Bottom: Same for a high latitude bin. 
The histograms are normalized to the sum of channels equals 1.}
\label{EW}
\end{figure}

Upper figure~\ref{EW} shows the experimental and simulated (with 1 $\sigma$ statistical 
error bars) ratios of the $e^+$ over $e^-$ energy-integrated yields as a function of the 
geomagnetic latitude. The agreement is good, within 1-2~$\sigma$, for all data points. 
A simulation run has been made with the $\pi^+$ production cross section taken equal to the 
$\pi^-$ one. In this case the asymmetry obtained is the same as for the normal calculation, 
to within statistical errors. The charge asymmetry of the $\pi$ production cross section 
then has a minor, if any, contribution to the observed lepton asymmetry.
\par
%
Now, let the EW angle be defined as the angle between particle momentum and meridian plane 
in geographical coordinates ($+\frac{\pi}{2}$ for E-bound particles). Lower 
figure~\ref{EW} shows the EW angle distributions of the lepton-producing particle momenta 
at the interaction point in the equatorial (upper) and high latitude (lower) regions (see 
latitude bin values on the figure). It shows very clearly that the $e^+$ and $e^-$ 
populations around the equator, are generated by (mostly proton) particle momenta quasi 
exclusively oriented eastward for $e^+$ and westward for $e^-$. The $e^+$ overall flux is  
more than 3 times larger than the $e^-$ flux, whereas at high latitudes the EW asymmetry is 
much weaker and the flux are about equal. 
This asymmetry originates from the well known GC dependence on the EW angle and results in 
the lepton charge asymmetry observed, because of the following specific conditions:
\begin{enumerate}
\item This EW angle dependence consists of a GC momentum P$_{GC}$ much lower for E-bound 
(P$^E_{GC}\approx$11~GeV/c) than for W-bound (P$^W_{GC}\approx$60~GeV/c) protons \cite{EW}. 
Consequently, a much larger flux of E-bound protons is allowed to interact in the atmosphere, 
since the incident flux decreases rapidly with the particle momentum as $E^{-2.7}$, E being 
the particle energy \cite{SI83}. 

\item The $\pi$ production cross section is peaked at forward angles, and this direction is 
approximately conserved through the decay process (for electrons/positrons above 0.2~GeV).  
The produced leptons then have about the same direction as incident (proton) particles. The 
proton flux asymmetry then translates into a lepton flux asymmetry through this direction
conserving effect. 

\item The acceptance of the earth-magnetosphere system is larger for E-bound than for W-bound 
secondary positrons, and conversely for electrons. Qualitatively, this is because any particle 
bent up by the earth magnetic field will propagate outside the atmosphere, whereas if bent 
down it will be absorbed with a high probability in the atmosphere. Therefore (E-bound) 
protons will produce a larger flux of (bent up) positrons.
\end{enumerate}

The above three conditions are required for the observed asymmetry to exist. For example,
a forward-backward symmetric lepton production would wash out the observed effect. Note that 
the sign of the asymmetry would change with the sign of the incident CR particles.
\par
This effect decreases progressively with the increasing latitudes because of the decreasing 
EW difference of P$_{GC}$. It is expected to vanish at the poles. 
\par
These results confirm the qualitative estimate from \cite{MING}, made on the basis of 
assumptions in agreement with our conclusions. See also \cite{LI01}.
\par
The contributions of the other CR components, $^4He$, $^{12}C$, $^{16}O$, etc.., should 
scale roughly with their respective flux, with some enhancement due to the larger total 
reaction cross sections however. It is then expected to be of the order of 10-20\% of the 
proton yield, i.e., within the fringe of uncertainty of the analysis. This is being evaluated 
quantitatively for $^4He$.
\par
In conclusion, it has been shown that the $e^+$ and $e^-$ populations measured by AMS below 
the geomagnetic cutoff can be well reproduced by a simulation assuming they originate from the 
hadronic production of $\pi^+$ and $\pi^-$ in proton collisions with atmospheric nuclei. The 
asymmetry of populations observed for $e^+$ and $e^-$ is due to a combination of: East-West 
asymmetry of the geomagnetic cutoff, forward peaking of the production cross section
and atmospheric absorption of the produced leptons. This conclusion could be reached on the 
basis of the good agreement between experimental data and simulation results. The difference 
between the $\pi^+$ and $\pi^-$ production cross sections plays a minor role in the 
phenomenon. \\
\par
$\underline{Acknowledgements}$: The authors are indebted to N. Chamel for his contribution 
to the first steps of this work. They also acknowledge discussions on the subject with 
F. Mayet, J. Favier, Haitao Liu, and members of the AMS collaboration. They are grateful to 
Dr V.V. Mikhailov for pointing out unnoticed references on the subject.

\begin{thebibliography}{99}
\bibitem{ALB} Treiman, Phys. Rev., 91, 957 (1953); Bland, Space Res., 5(1965)618; 
              S.D. Verma, J. Geophys. Res., 72(1967)915;
              M.H. Israel, J. Geophys. Res., 74(1969)4701; 
              Konig, J. Geophys. Res., 86(1981)515

\bibitem{PROT} The AMS collaboration, J. Alcaraz et al., Phys. Lett. B472(2000)215
\bibitem{LEPT} The AMS collaboration, J. Alcaraz et al., Phys. Lett. B484(2000)10
\bibitem{NEUTRINOS} see for example: 
               R. Engel, T.K. Gaisser, and T. Stanev, Phys. Lett. B472(2000)113;
               T.K. Gaisser et al., Phys. Rev. D54(1996)5578
\bibitem{PAP1} L.Derome et al., Phys. Lett. B 489(2000)1
\bibitem{SI83} J.A. Simpson, Ann. Rev. Nucl. and Part. Sci. 33(1983)323
\bibitem{EW}   See for exampe, M. Longair, High Energy Astrophysics, Cambridge UP, 1994;
               A.E. Sandstr\"om, Cosmic Ray Physics, NHPC, 1965, Chapter 5;
\bibitem{COCH} D.R.F. Cochran et al., Phys. Rev. D6(1972)3085; 
\bibitem{PICS} M. Antinucci et al., Let. Nuov. Cim. 6(1973)121.
               H. Machner and J. Haidenbauer, J. Phys. G 25(1999)R231
\bibitem{GR77} N.L. Grigorov, Sov. Phys. Dokl., 22(1977)305
\bibitem{VO94} S.A. Voronov, S.V. Koldashov, and V.V. Mikhailov, Cosmic Research, 
               33(95)300
\bibitem{COSP} The AMS collaboration, J. Alcaraz et al., Phys. Lett. B490(2000)27
\bibitem{CHO}   Y. Cho et al., Phys. Rev. D4(1971)1967
\bibitem{ABBO}  T. Abbott et al., Phys. Rev. D45(1992)3906
\bibitem{PIDAT} J. Papp, thesis, LBL report 3633, Berkeley, 1975;  
                T. Eichten et al., Nucl. Phys. B44(1972)333;
                V.V. Abramov et al., Z. Phys. C24(1984)205;
                D. Antreasyan et al. Phys. Rev. D19(1979)764;
                H. B{\o}ggild et al., Phys. Rev. C59(1999)328
\bibitem{LPIDAT} J. Martensson et al., Phys. Rev. C62(2000)014610; 
                 J.F.Crawford et al., Phys. Rev. C22(1980)1184

\bibitem{KALI} A.M. Kalinovski, N.V. Mokhov, and Yu.P. Nikitin, {\it Passage of 
               high energy particles through matter,} AIP edit, New-York, 1989, p 101
\bibitem{MOKH} N.V. Mokhov and S.I. Striganov, Proc. of Worksh. on Phys. at First 
               Muon Collider, Fermilab, Nov 6-9, 1997; 
               and N.V. Mokhov, and private communication. 
\bibitem{YONG} Y. Liu, L. Derome and M. Buénerd, internal report ISN-01-12
\bibitem{PI0} S.R. Blattnig et al., Phys. Rev. D62:0940030,2000
\bibitem{MING} M-H. A. Huang, Proc. of the 8th Asian Pacific Physics conf.,
               Aug. 7-10, Taipei (Taiwan), World Scientific, to appear.
\bibitem{LI01} P. Lipari, astro-ph/0101559, jan 31, 2001.
%
\end{thebibliography}
\end{document}